\documentclass[fleqn,twoside]{article}
\newcommand{\lesssim}{\raisebox{0.3mm}{\em $\, <$} \hspace{-2.8mm}
\raisebox{-1.3mm}{\em $\sim \,$}}
\newcommand{\gtrsim}{\raisebox{0.3mm}{\em $\, >$} \hspace{-2.8mm}
\raisebox{-1.3mm}{\em $\sim \,$}}
\usepackage{espcrc2}

\usepackage{graphicx}
\usepackage[figuresright]{rotating}

\newcommand{\AmS}{{\protect\the\textfont2
  A\kern-.1667em\lower.5ex\hbox{M}\kern-.125emS}}

\title{Degeneracy and strategies of long baseline and reactor
experiments
\thanks{Work supported in part by Grants-in-Aid for Scientific Research
No.\ 16540260 and No.\ 16340078, Japan Ministry
of Education, Culture, Sports, Science, and Technology.}
}

\author{Osamu Yasuda\\
        {\ }\\
        Department of Physics, Tokyo Metropolitan University\\
        1-1 Minami-Osawa Hachioji, Tokyo 192-0397, Japan}
       
\begin{document}

\begin{abstract}
Assuming that the JPARC
experiment measures the oscillation probabilities
$P(\nu_\mu\rightarrow\nu_e)$ and
$P(\bar{\nu}_\mu\rightarrow\bar{\nu}_e)$ at $|\Delta m^2_{31}|L/4E=\pi/2$,
I discuss what kind of extra experiment (long baseline or reactor)
can contribute to determination of $\theta_{13}$
and the CP phase $\delta$.
\vspace{1pc}
\end{abstract}

\maketitle

\section{Introduction}

Determination of the unknown oscillation parameters
$\theta_{13}$ and $\delta$ is the important object of
future neutrino experiments.
It has been
known that even if the values of the oscillation
probabilities $P(\nu_\mu \rightarrow \nu_e)$ and
$P(\bar{\nu}_\mu \rightarrow \bar{\nu}_e)$ are exactly given
we cannot determine uniquely the values of the
oscillation parameters due to three kinds of parameter degeneracies:
the intrinsic $(\theta_{13}, \delta)$ 
degeneracy,
the degeneracy of
$\Delta m^2_{31}\leftrightarrow-\Delta m^2_{31}$,
and the degeneracy of
$\theta_{23}\leftrightarrow\pi/2
-\theta_{23}$.
Each degeneracy gives a twofold ambiguity, so
in general we have an eightfold ambiguity.
These degeneracies cause a problem in determination of
$\theta_{13}$ and $\delta$, and we have to take into account
these ambiguities in handling
the data in future long baseline experiments.
Here I will discuss only the oscillation probabilities
without details analysis of statistical and systematic errors.

\section{Determination of $\sin^22\theta_{13}$}

There is a way to overcome the ambiguity due to the intrinsic
degeneracy.  Namely, if one performs a long baseline experiment at the
oscillation maximum (i.e., with $|\Delta m^2_{31}L/4E|=\pi/2$), then
it is reduced to the $\delta\leftrightarrow\pi-\delta$ ambiguity.

Thus, if the JPARC experiment measures the oscillation probabilities
$P(\nu_\mu\rightarrow\nu_e)$,
$P(\bar{\nu}_\mu\rightarrow\bar{\nu}_e)$ and
$P(\nu_\mu\rightarrow\nu_\mu)$ at the oscillation maximum
with an approximately monoenergetic beam, then the possible solutions
in the ($\sin^22\theta_{13}$, $1/s^2_{23}$) plane \cite{Yasuda:2004gu} given by
JPARC look like either Fig. 1(a)
or Fig. 1(b), depending on whether $\cos^22\theta_{23}\ll{\cal O}(0.1)$
or $\cos^22\theta_{23}\simeq{\cal O}(0.1)$, where the difference
between the case with normal hierarchy ($\Delta m^2_{31}>0$) and
the one with inverted hierarchy ($\Delta m^2_{31}<0$) is small
because the matter effect is small in the JPARC experiment.

\vspace{-10pt}
\begin{center}
\begin{figure}[htb]
\vspace{-10pt}
\hglue -0.5cm
\includegraphics[width=45mm]{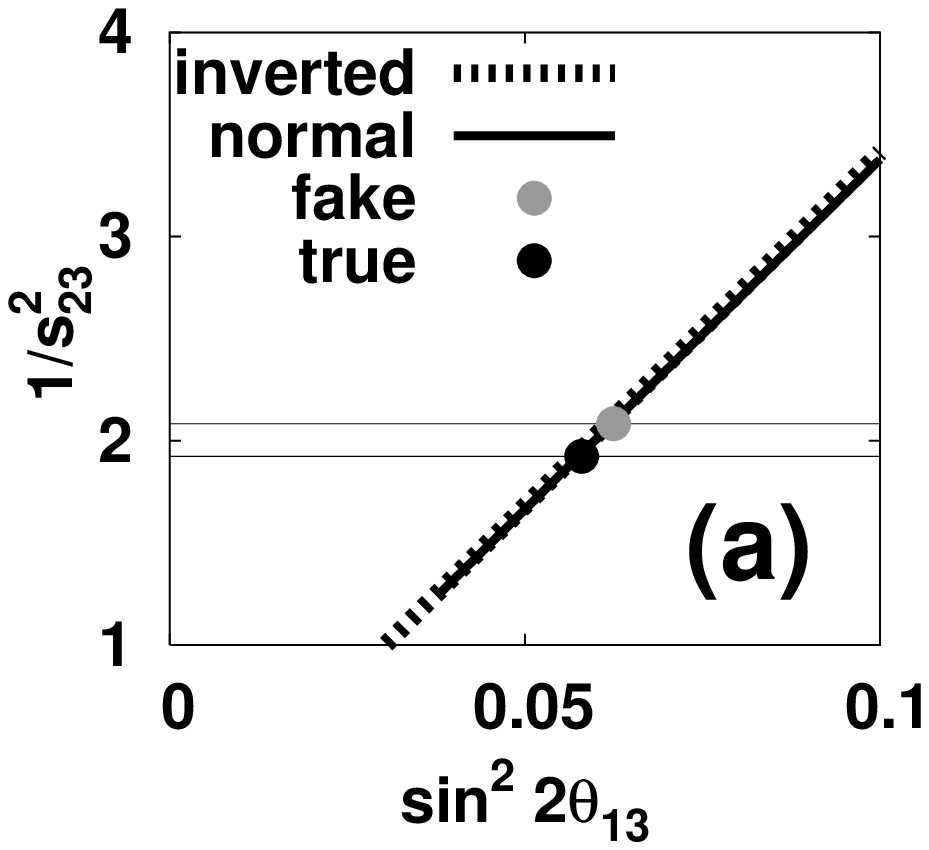}
\vglue -4.2cm\hglue 3.5cm
\includegraphics[width=45mm]{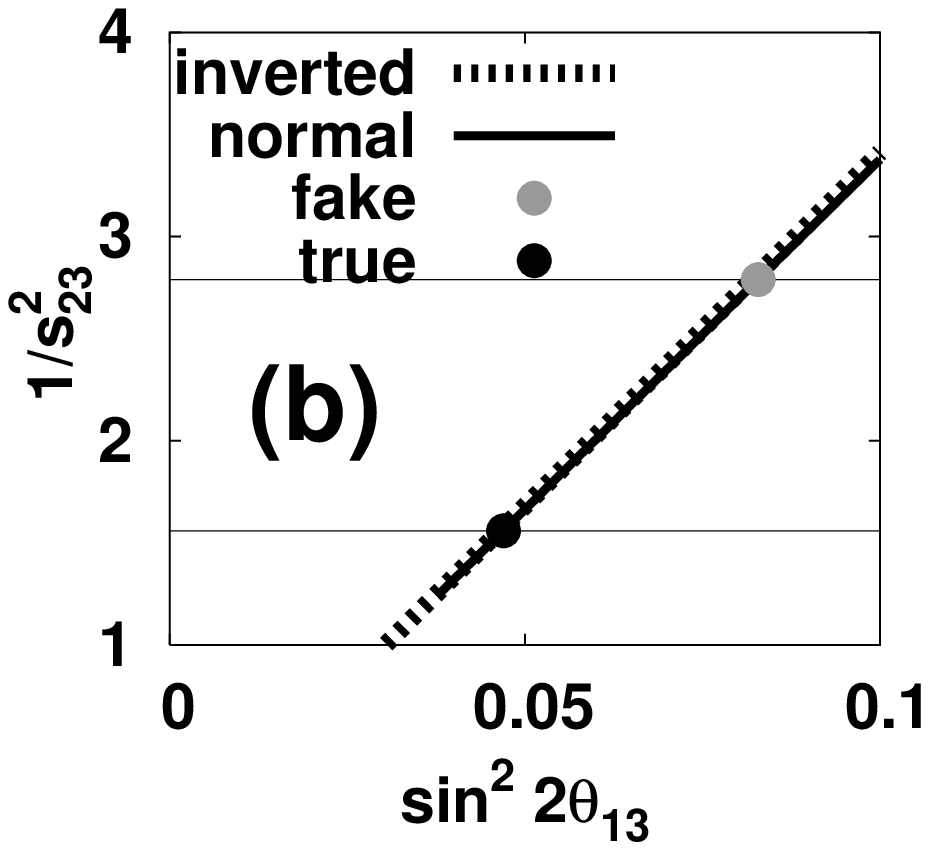}
\vspace{-40pt}
\caption{The $\theta_{23}$ ambiguity which could
arise after the JPARC measurements of
$P(\nu_\mu\rightarrow\nu_e)$,
$P(\bar{\nu}_\mu\rightarrow\bar{\nu}_e)$ and
$P(\nu_\mu\rightarrow\nu_\mu)$ at the oscillation
maximum.
(a) The case of $\cos^22\theta_{23}\ll{\cal O}(0.1)$.
(b) The case of $\cos^22\theta_{23}\simeq{\cal O}(0.1)$.
}
\label{fig1}
\end{figure}
\end{center}
\vspace{-30pt}

If $\cos^22\theta_{23}\ll{\cal O}(0.1)$ then
the values of $\theta_{13}$ and $\theta_{23}$
are close to each other for all the solutions, and the
ambiguity is not serious as far as $\theta_{13}$ and $\theta_{23}$ are
concerned (cf. Fig.\ref{fig1}(a)).
On the other hand, if $\cos^22\theta_{23}\simeq{\cal O}(0.1)$ then
the $\theta_{23}$ ambiguity has to be resolved to determine
$\theta_{13}$ and $\theta_{23}$ to good precision  (cf. Fig.\ref{fig1}(b)).
In this case there are three potential possibilities to resolve
this $\theta_{23}$ ambiguity:
(a) reactor experiments, (b) the $\nu_\mu \to \nu_e$ (or $\nu_e \to \nu_\mu$)
channel of accelerator long baseline experiments
and (c) the so-called silver channel $\nu_e \to \nu_\tau$.

\begin{center}
\begin{figure}[ht]
\vspace{-20pt}
\hglue -0.7cm
\includegraphics[width=47mm]{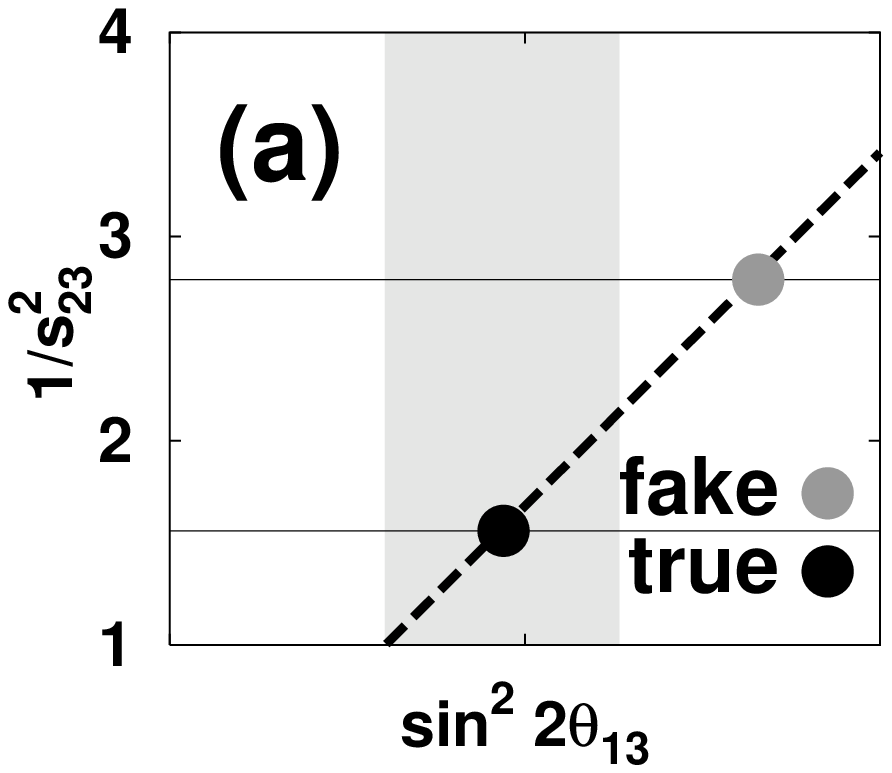}
\vglue -4.5cm\hglue 3.5cm
\hglue 0.1cm
\includegraphics[width=45mm]{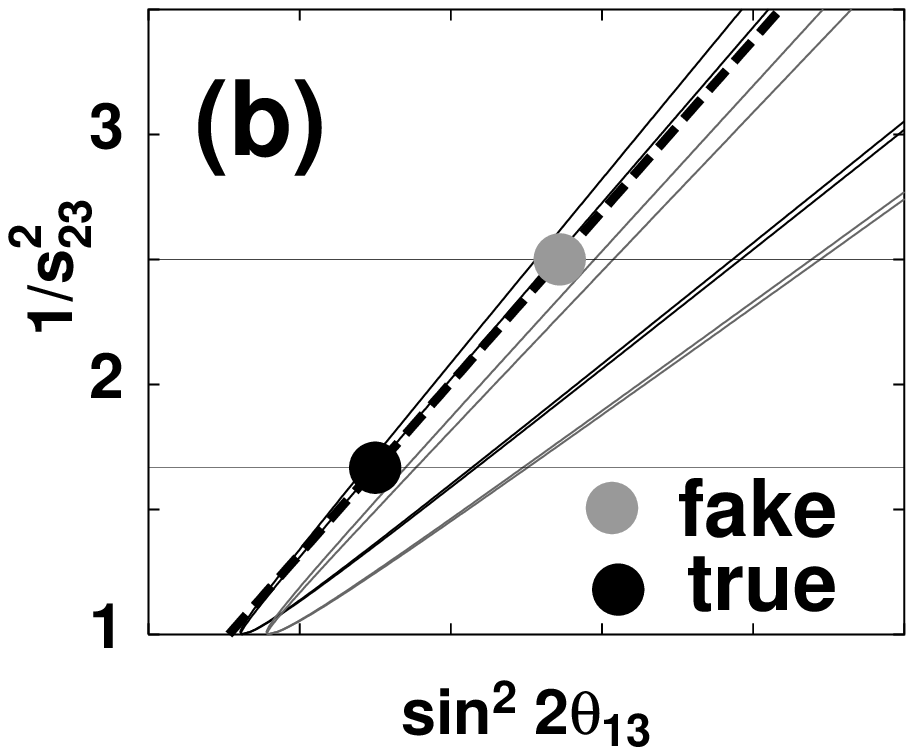}
\vglue -0.3cm\hglue -0.5cm
\includegraphics[width=42mm]{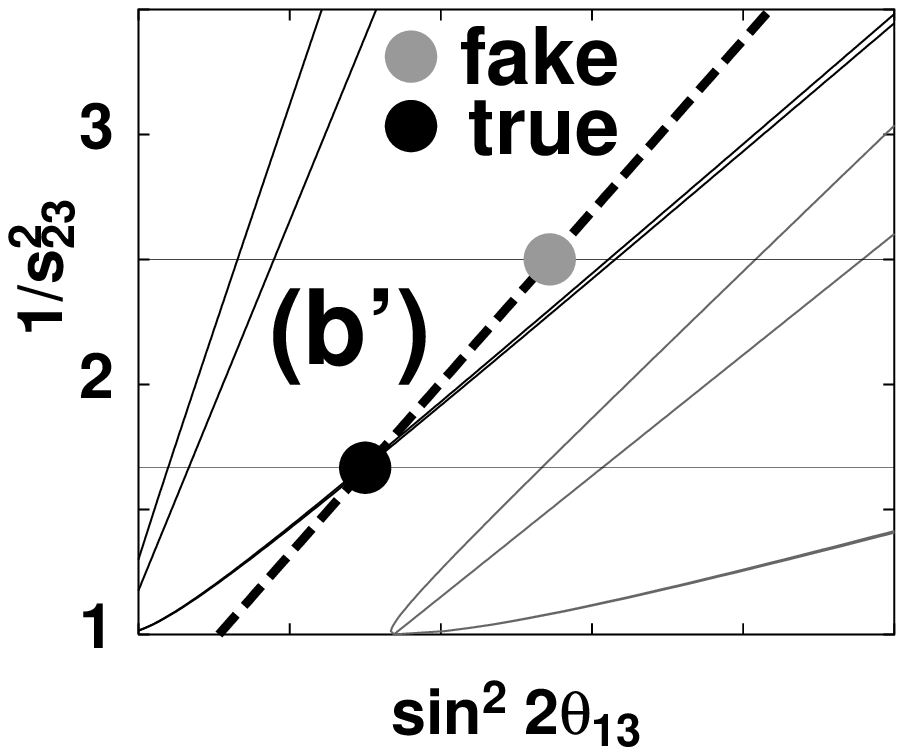}
\vglue -3.8cm\hglue 3.8cm
\includegraphics[width=47mm]{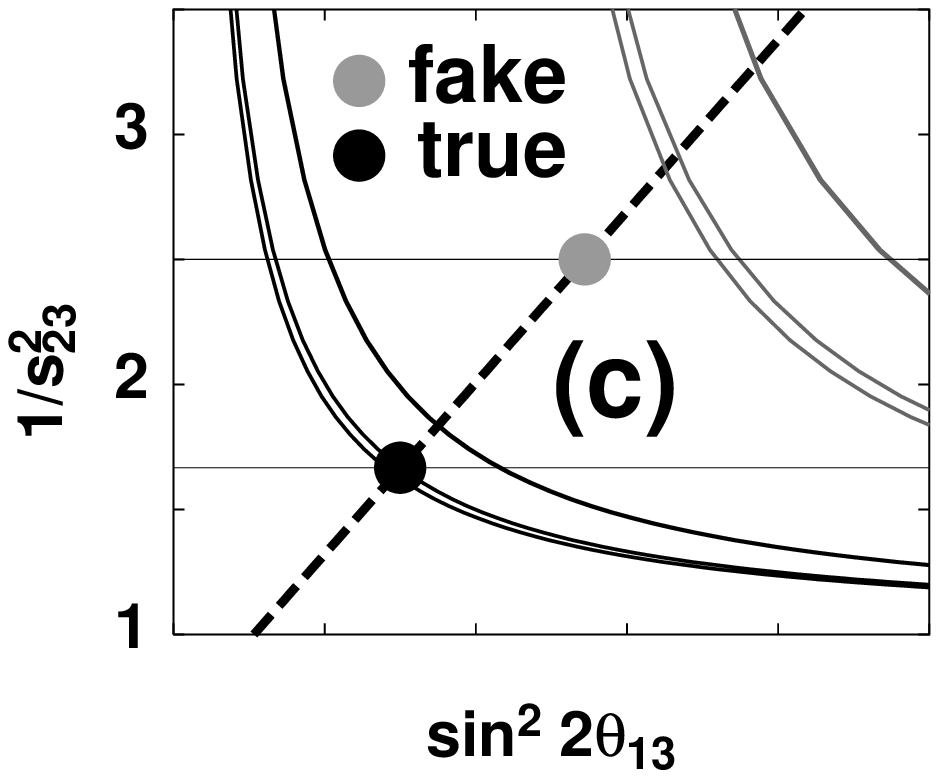}
\vspace{-40pt}
\caption{
(a) The case of a reactor experiment,
which gives the constraint only on  $\sin^22\theta_{13}$
(the shadowed region).
(b,b') The case of the $\nu_\mu \to \nu_e$ (or $\nu_e \to \nu_\mu$)
channel with the baseline $L$=730km and the neutrino energy $E$=6GeV ((b)),
$E$=1GeV ((b')).
(c) The case of the $\nu_e \to \nu_\tau$ channel with $L$=3000km and
$E$=12GeV.  The dashed line is
the constraint by the JPARC measurements of
$P(\nu_\mu\rightarrow\nu_e)$ and $P(\bar{\nu}_\mu\rightarrow\bar{\nu}_e)$,
where the two hierarchical patterns
are identified for simplicity because the difference between them is very small,
and the thin black and gray curves
stand for those by the third experiment with correct and wrong
assumptions on the mass hierarchy.
See \cite{Yasuda:2004gu} for details of the figures.
}
\label{fig2}
\end{figure}
\end{center}

A reactor experiment measures
$P(\bar{\nu}_e\rightarrow\bar{\nu}_e)$ which depends only
on $\theta_{13}$ to a good approximation, so that it
gives a constraint as a vertical band in the 
($\sin^22\theta_{13}$, $1/s^2_{23}$) plane (cf. Fig.\ref{fig2}(a)).
Thus, if the experimental error in the reactor experiment
is smaller than the difference between the true and fake values
of $\theta_{13}$, then
the reactor experiment can solve the $\theta_{23}$ ambiguity.
To resolve the $\theta_{23}$ ambiguity at a high confidence level,
it is necessary for a reactor experiment to have high
sensitivity (e.g., $\left(\sin^22\theta_{13}
\right)_{\mbox{\rm\scriptsize sensitivity}}\lesssim0.01$).
It is known \cite{Sugiyama:2004bv}
that the sensitivity in reactor experiments
is bounded from below by the uncorrelated systematic error
$\sigma_{\mbox{\rm\scriptsize u}}$ of the detector:
\begin{eqnarray}
\left(\sin^22\theta_{13}\right)_{\mbox{\rm\scriptsize sensitivity}}
\ge \mbox{\rm const.}\sigma_{\mbox{\rm\scriptsize u}},
\label{sensitivity}
\end{eqnarray}
where const. depends on the numbers of the reactors and the detectors,
and is equal to 2.8 at 90\%CL in the case with one reactor and two detectors.
Eq. (\ref{sensitivity}) implies that it is necessary
to reduce $\sigma_{\mbox{\rm\scriptsize u}}$
in order to improve the sensitivity.
If the uncorrelated
systematic error $\sigma_{\mbox{\rm\scriptsize u}}$ of the detectors
is independent of the number of the detectors,
then there is a way to reduce $\sigma_{\mbox{\rm\scriptsize u}}$.
In the case with one reactor,
if one puts $N$ identical detectors at the near site
and $N$ identical detectors at the far site, where all these detectors
are assumed to have the same uncorrelated systematic error
$\sigma_{\mbox{\rm\scriptsize u}}$, then the lower bound of
the sensitivity becomes
\begin{eqnarray}
\mbox{\rm lower bound of }
\left(\sin^22\theta_{13}\right)_{\mbox{\rm\scriptsize sensitivity}}
={2.8 \over \sqrt{N}} \,\sigma_{\mbox{\rm\scriptsize u}}.
\nonumber
\end{eqnarray}
Hence it follows theoretically that the more identical detectors one
puts, the better sensitivity one gets.  The assumption that
$\sigma_{\mbox{\rm\scriptsize u}}$ is independent of $N$ may not be
satisfied in reality, but if the dependence of
$\sigma_{\mbox{\rm\scriptsize u}}$ on $N$ is weaker than $\sqrt{N}$,
$\sigma_{\mbox{\rm\scriptsize u}}/\sqrt{N}$ decreases as $N$
increases.  This possibility should be seriously thought about to
improve the sensitivity in the future reactor experiments.

The $\nu_\mu \to \nu_e$ (or $\nu_e \to \nu_\mu$)
channel gives poor resolution in general, as far as
the $\theta_{23}$ ambiguity is concerned,
because the curves given by this channel
intersect almost in parallel with the JPARC line and
the experimental significance to reject the wrong hypothesis
is expected to be small (cf. Fig.\ref{fig2}(b),(b')).
The situation changes very little even if one uses
$\bar{\nu}_\mu \to \bar{\nu}_e$ (or $\bar{\nu}_e \to \bar{\nu}_\mu$).

On the other hand, the silver channel $\nu_e \to \nu_\tau$
offers a promising possibility to resolve the $\theta_{23}$
ambiguity, because the curves given by the silver channel
intersect with the JPARC line almost perpendicularly and
the experimental significance to reject the wrong hypothesis
is expected to be large (cf. Fig.\ref{fig2}(c)).

\vspace{-35pt}
\section{Determination of $\delta$}
\begin{center}
\begin{figure}[htb]
\vspace{-20pt}
\begin{center}
\includegraphics[width=70mm]{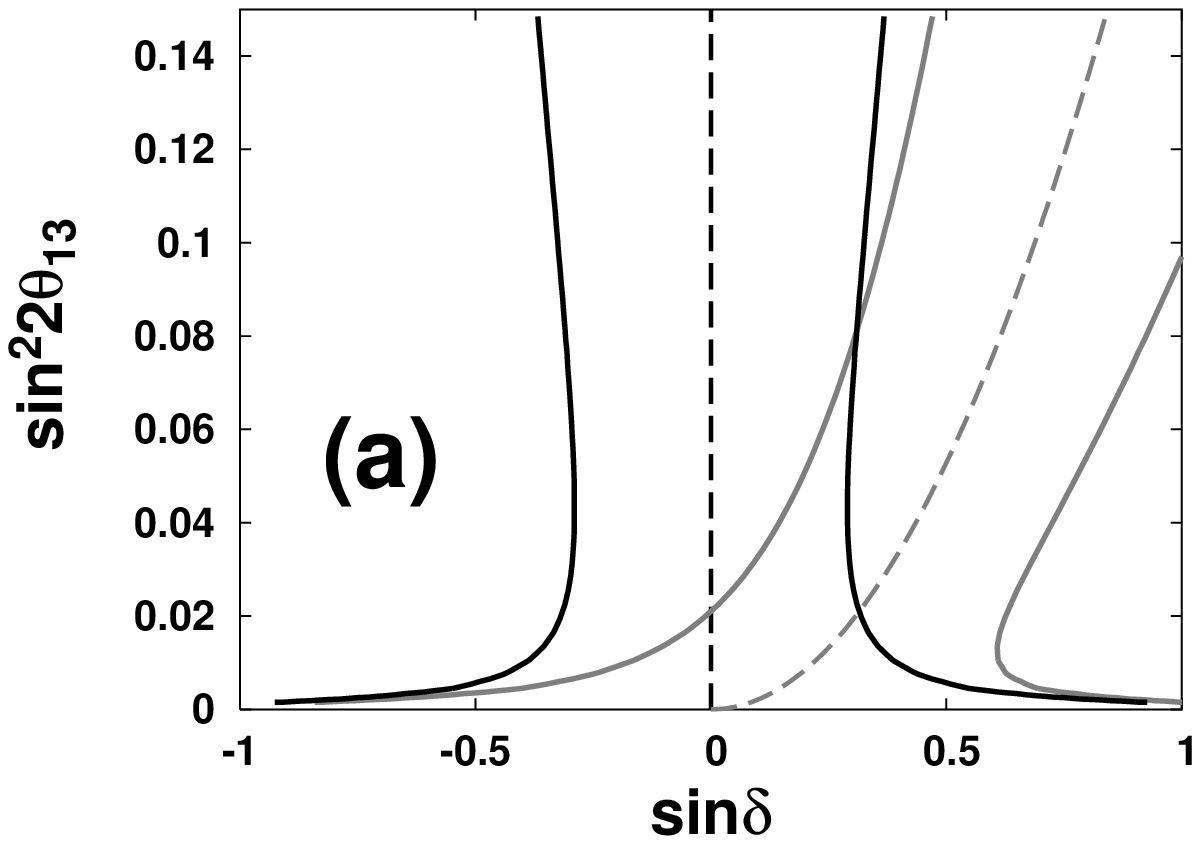}
\vglue -0.5cm
\includegraphics[width=70mm]{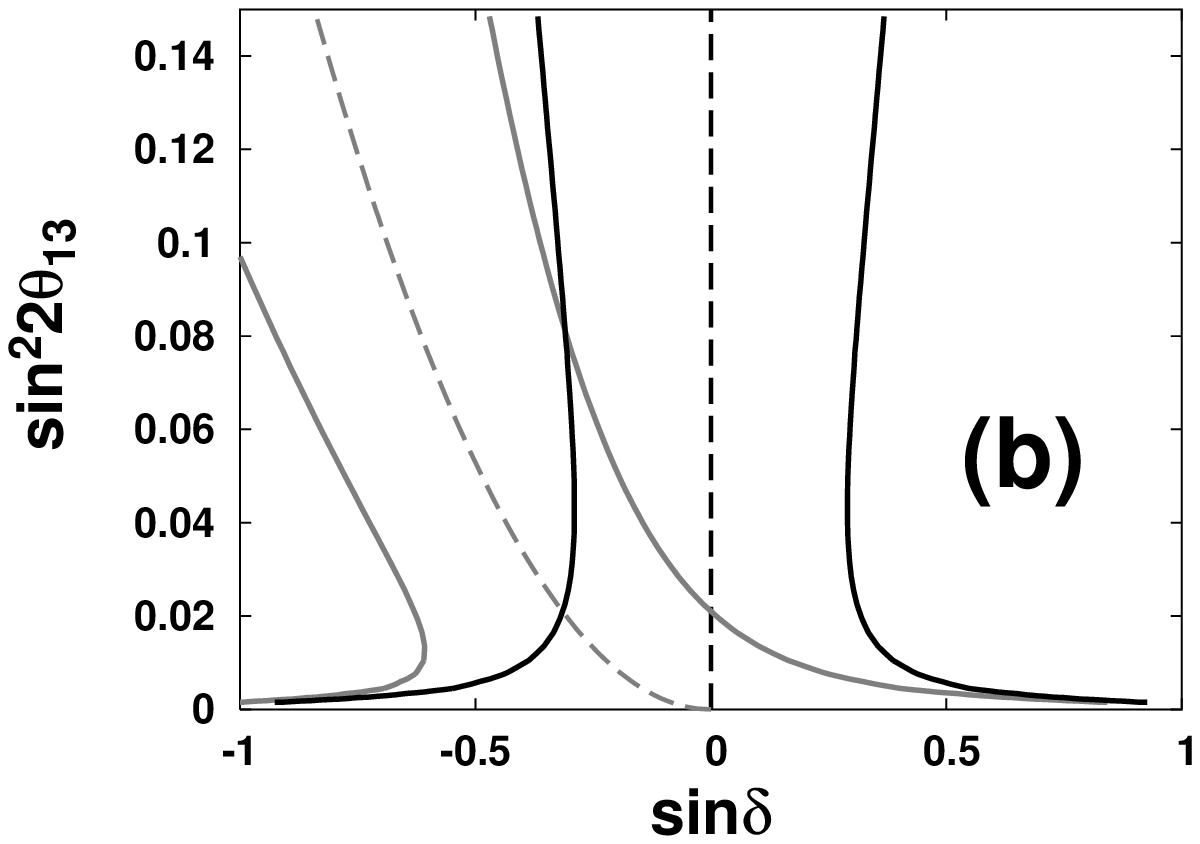}
\end{center}
\vspace{-30pt}
\caption{
The sensitivity to CP violation
at 3$\sigma$ in the case of the JPARC experiment.  The black (grey) curves give
the sensitivity to CP violation with the correct (wrong)
assumption on the mass hierarchy when $\Delta m^2_{31}>0$
(a) or when $\Delta m^2_{31}<0$ (b).
}
\label{fig3}
\end{figure}
\end{center}

If the JPARC experiments for $\nu_\mu \rightarrow \nu_e$
and $\bar{\nu}_\mu \rightarrow \bar{\nu}_e$
are performed at the oscillation maximum,
then the value of $\sin\delta$ suffers from the two
ambiguities, i.e., those due to sgn($\Delta m^2_{31}$)
and $\theta_{23}\leftrightarrow\pi/2-\theta_{23}$,
since the $\delta\leftrightarrow\pi-\delta$ ambiguity
does not affect the value of $\sin\delta$.
To simplify the discussions, let me discuss the case where
the reference value of $\delta$ is zero.  In this case
the fake values
$\left.\sin\delta'\right|_{\mbox{\rm\scriptsize sgn}(\Delta m^2_{31})}$
due to the sgn($\Delta m^2_{31}$) ambiguity and
$\left.\sin\delta'\right|_{\mbox{\rm\scriptsize sgn}(\cos2\theta_{23})}$
due to the
$\theta_{23}\leftrightarrow\pi/2-\theta_{23}$ ambiguity
satisfy the following for the JPARC case \cite{Yasuda:2004tw}:
\begin{eqnarray}
\hspace*{7mm}
\left.\sin\delta^\prime\right|_{\mbox{\rm\scriptsize sgn}(\Delta m^2_{31})}
&\simeq&-2.2\sin\theta_{13},\nonumber\\
\left|
\left.\sin\delta'\right|_{\mbox{\rm\scriptsize sgn}(\cos2\theta_{23})}
\right|
&\lesssim&
\frac{1}{500}\frac{1}{\sqrt{\sin^22\theta_{13}}}.\nonumber
\end{eqnarray}
This suggests that the $\theta_{23}$ ambiguity does not cause
a serious problem for $\sin^22\theta_{13}\gtrsim10^{-3}$,
while the one due to the sgn($\Delta m^2_{31}$) ambiguity does.
In Fig.\ref{fig3} the sensitivity to CP violation is given
at 3$\sigma$ by a semi-quantitative analysis for the JPARC experiment
in the case of
(a) $\Delta m^2_{31}>0$ and (b) $\Delta m^2_{31}<0$,
taking into account of the ambiguity due to
sgn($\Delta m^2_{31}$).
The the dashed lines in the figures
are given by $\sin\delta=0$ for the correct assumption, and
by $\sin\delta=\pm2.2\sin2\theta_{13}$ for the wrong
assumption.
In the region bounded by the
black (with the correct assumption on the mass hierarchy)
or grey (with the wrong assumption)
curves, JPARC cannot claim CP violation to be nonzero.
From this, therefore, we see that it is important to
resolve the sgn($\Delta m^2_{31}$) ambiguity to determine
the precise value of $\delta$.

To resolve the sgn($\Delta m^2_{31}$) ambiguity, it is easy to see
that long baseline experiments with long baselines ($\gtrsim$1000km)
are advantageous, since the matter effect $\sqrt{2}G_FN_e$ for the
density $\rho\simeq3g/cm^3$ is something like 1/2000km.  What is not
trivial to see is the fact that a long baseline experiment with {\it lower}
neutrino energy is advantageous for the {\it same} baseline $L$, as
long as the energy $E$ satisfies
$\left|\Delta m^2_{31}L/4E\right|<\pi$ \cite{Yasuda:2004gu}.
Hence, although the NOvA experiment has a baseline
shorter than 1000km, if it is performed with lower energy
(e.g., 1GeV), then it may have a chance to resolve
the sgn($\Delta m^2_{31}$) ambiguity.

\section{Summary}
If $\cos^22\theta_{23}\simeq{\cal O}(0.1)$ then
it is important to resolve the $\theta_{23}$ ambiguity
to determine the value of $\theta_{13}$ by an experiment
other than JPARC.
On the other hand, to determine the value of $\delta$,
it is important to resolve the sgn($\Delta m^2_{31}$) ambiguity
irrespective of the value of $\theta_{13}$.

\end{document}